\documentclass[11pt]{article}

\usepackage[preprint]{acl}

\usepackage{times}
\usepackage{latexsym}
\usepackage{graphicx}
\usepackage[T1]{fontenc}

\usepackage[utf8]{inputenc}

\usepackage{microtype}
\usepackage{amsmath}
\usepackage{amssymb}
\usepackage{multirow} 
\usepackage{booktabs} 
\usepackage{colortbl}  
\usepackage{inconsolata}

\usepackage{graphicx}

\title{SAFE-QAQ: End-to-End Slow-Thinking Audio-Text Fraud Detection via Reinforcement Learning}

\author{
    Peidong Wang\textsuperscript{1}\thanks{\ \ These authors contributed equally to this research.} \quad
    Zhiming Ma\textsuperscript{1,2}\footnotemark[1] \quad
    Xin Dai\textsuperscript{1}\footnotemark[1] \quad
    Yongkang Liu\textsuperscript{1} \quad
    \textbf{Shi Feng}\textsuperscript{1}\thanks{\ \ Corresponding author.} \\
    \textbf{Xiaocui Yang}\textsuperscript{1} \quad
    \textbf{Wenxing Hu}\textsuperscript{3} \quad
    \textbf{Zhihao Wang}\textsuperscript{1} \quad
    \textbf{Mingjun Pan}\textsuperscript{2} \quad
    \textbf{Li Yuan}\textsuperscript{4} \quad
    \textbf{Daling Wang}\textsuperscript{1} \\
    \textsuperscript{1}Northeastern University, China \\
    \textsuperscript{2}China Mobile Internet Company Ltd. \\
    \textsuperscript{3}Shanghai University of Electric Power, China \\
    \textsuperscript{4}Peking University, Shenzhen, China \\
    \texttt{pdongwang@163.com}, \texttt{mazhiming312@outlook.com} \\
    \texttt{daix1@mails.neu.edu.cn}, 
    \texttt{fengshi@cse.neu.edu.cn}
}

\begin{document}
\maketitle

\begin{abstract}
  Existing fraud detection methods predominantly rely on transcribed text, suffering from ASR errors and missing crucial acoustic cues like vocal tone and environmental context. This limits their effectiveness against complex deceptive strategies.
  To address these challenges, we first propose \textbf{SAFE-QAQ}, an end-to-end comprehensive framework for audio-based slow-thinking fraud detection. First, the SAFE-QAQ framework eliminates the impact of transcription errors on detection performance. Secondly, we propose rule-based slow-thinking reward mechanisms that systematically guide the system to identify fraud-indicative patterns by accurately capturing fine-grained audio details, through hierarchical reasoning processes. Besides, our framework introduces a dynamic risk assessment framework during live calls, enabling early detection and prevention of fraud. Experiments on the TeleAntiFraud-Bench demonstrate that SAFE-QAQ achieves dramatic improvements over existing methods in multiple key dimensions, including accuracy, inference efficiency, and real-time processing capabilities. Currently deployed and analyzing over 70,000 calls daily, SAFE-QAQ effectively automates complex fraud detection, reducing human workload and financial losses. Code: \url{https://anonymous.4open.science/r/SAFE-QAQ}.
\end{abstract}



\section{Introduction}

\begin{figure}[t]
    \centering
    \includegraphics[width=\linewidth]{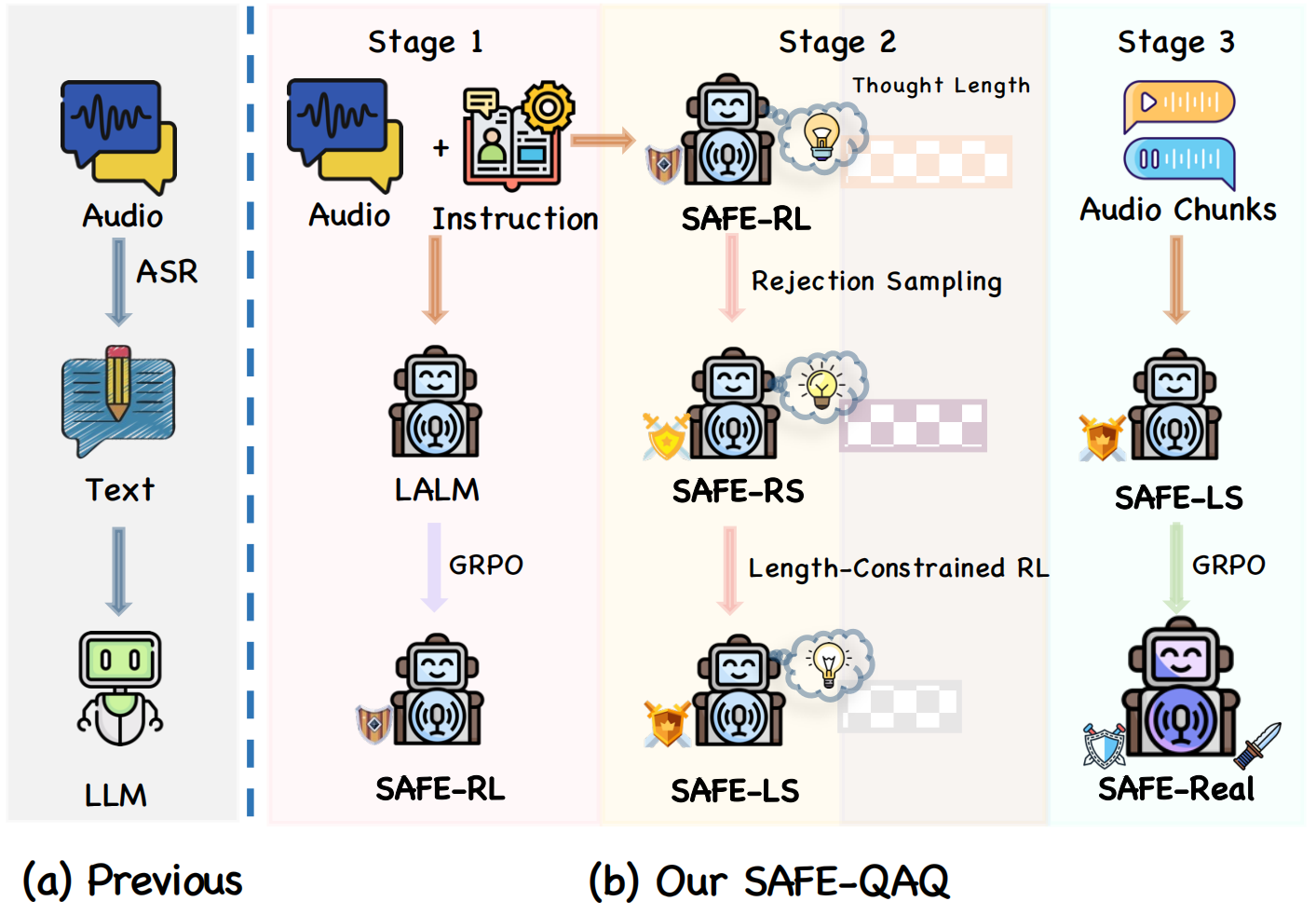}
    \caption{Comparison of (a) Previous Method and (b) Our Proposed Method: End-to-End Call Fraud Detection via Reinforcement Learning (RL). Our approach trains LALMs in three stages: (i) developing slow-thinking reasoning through RL, (ii) optimizing thought length using rejection sampling fine-tuning and length-constrained RL, and (iii) achieving real-time detection with audio chunks training.}
    \label{fig:intro}
\end{figure}
With the rapid development of mobile communication technology, the problem of telecom fraud has become increasingly severe, emerging as a global social challenge. 
As illustrated in Figure \ref{fig:intro}(a), most existing methods \cite{chang2024exposing, korkanti2024enhancing, yang2025fraud, shen2025warned, hu2024zipzap} for fraud detection have made significant advancements by utilizing transcribed texts and the powerful representation capabilities of large language models (LLMs) \cite{dubey2024llama, achiam2023gpt, guo2025deepseek}. However, these methods often result in misjudgments in real-time fraud detection systems due to their inability to update information and policies in a timely manner. To address this issue, researchers have proposed using Retrieval-Augmented Generation (RAG) to stay updated with the latest information and policies, eliminating the need for retraining \cite{singh2025advanced}. Despite these efforts, the application of fraud detection systems in real-world scenarios remains challenging and far from fully successful, primarily due to the following three reasons. 
First, the transcription of speech into text by Automatic Speech Recognition (ASR) systems can introduce error noise, leading to a decrease in model performance \cite{ma2025teleantifraud,hu2024zipzap, chakraborty2024detoxbench}. Second, text-based fraud detection systems cannot capture the fine-grained information conveyed by speech, such as vocal tone, emotional stress, and environmental acoustics, all of which are crucial for effective fraud detection. Third, modern fraudsters employ elaborate layered deceptive strategies—combining manipulated speech patterns (e.g., voice spoofing or synthetic audio), fabricated background sounds (e.g., fake call center noise), and psychologically coercive scripts—that require iterative reasoning to unravel \cite{ma2025teleantifraud}. However, current text-based pipeline methods lack mechanisms for deep reasoning to effectively address such complexities. These limitations highlight the inadequacy of text-only approaches in tackling modern telecommunications fraud and sophisticated deceptive strategies.

To address these challenges, we propose SAFE-QAQ (\textbf{S}low-thinking \textbf{A}udio-text \textbf{F}raud d\textbf{E}tection using \textbf{Q}wen \textbf{A}udio with \textbf{Q}uestion), an end-to-end comprehensive framework for audio-based slow-thinking fraud detection. Building upon recent advancements in Large Audio-Language Models (LALMs) \cite{chu2024qwen2, huang2025step, hurst2024gpt, zeng2024glm}, SAFE-QAQ establishes a complete end-to-end pipeline that directly processes raw audio signals to preserve crucial multimodal features while incorporating three key innovations (as shown in Figure \ref{fig:intro}(b)): 
\begin{itemize}
    \item SAFE-QAQ develops slow-thinking reasoning through rule-based reward, enabling the system to systematically analyze fine-grained details, which are frequently concealed by fraudsters using layered deceptive strategies. 
    \item we further optimize reasoning efficiency by reducing reasoning chain lengths by 48.87\% through rejection sampling fine-tuning (producing \emph{SAFE-RS}) and length-constrained RL (resulting in \emph{SAFE-LS}), ensuring concise yet accurate reasoning. 
    \item SAFE-QAQ achieves real-time detection by dynamically assessing information sufficiency through structured prompting and phase recognition rewards, culminating in the final model (\emph{SAFE-Real}) that enables timely interventions during live calls. 
\end{itemize}
By eliminating reliance on error-prone ASR transcriptions and integrating slow-thinking reasoning with reinforcement learning-optimized multimodal processing, SAFE-QAQ establishes a fully end-to-end framework that achieves both high accuracy and practical efficiency. This viability is demonstrated by its successful deployment in a production pipeline processing over 70,000 calls daily, where it effectively alleviates manual audit burdens and prevents financial losses through timely, automated intervention.

\section{Related work}



\subsection{LLM-Based Telecom Fraud Detection}

Recent advances in LLMs have shown promise for telecom fraud detection, with methods like Retrieval-Augmented Generation (RAG) for real-time call analysis \cite{singh2025advanced} and intent-based warning systems \cite{shen2025warned}. However, these approaches rely solely on transcribed text, discarding critical audio features (e.g., tone, emotion) that signal fraud \cite{chang2024exposing}. While multimodal benchmarks like TeleAntiFraud-28k \cite{ma2025teleantifraud} address this gap, their supervised fine-tuning (SFT) methods underutilize modern LLMs' reasoning capabilities. Current systems also suffer from inefficiency, requiring multi-stage pipelines for transcription and fraud detection \cite{yang2025fraud}. SAFE-QAQ overcomes these limitations by processing raw audio end-to-end via reinforcement learning (RL), preserving multimodal cues while eliminating intermediate steps. This approach enables faster, more accurate fraud detection tailored to real-world dynamics.

\subsection{Large Audio Language Models (LALMs)}

LALMs such as Qwen2-Audio, GLM-4-Voice, GPT-4o, and Step-Audio have shown strong performance in speech understanding, capturing tone, emotion, and intent in real time \cite{chu2024qwen2, zeng2024glm, hurst2024gpt, huang2025step}, but their application to fraud detection remains limited. General-purpose LALMs often fail to detect scripted deception, where vocal delivery subtly contradicts scripted calmness. \textbf{SAFE-QAQ} bridges this gap by aligning audio-language modeling with domain-specific RL optimization for telecom fraud detection, enabling context-aware, risk-sensitive reasoning.

\begin{figure*}[th]
    \centering
    \includegraphics[width=1.0\linewidth]{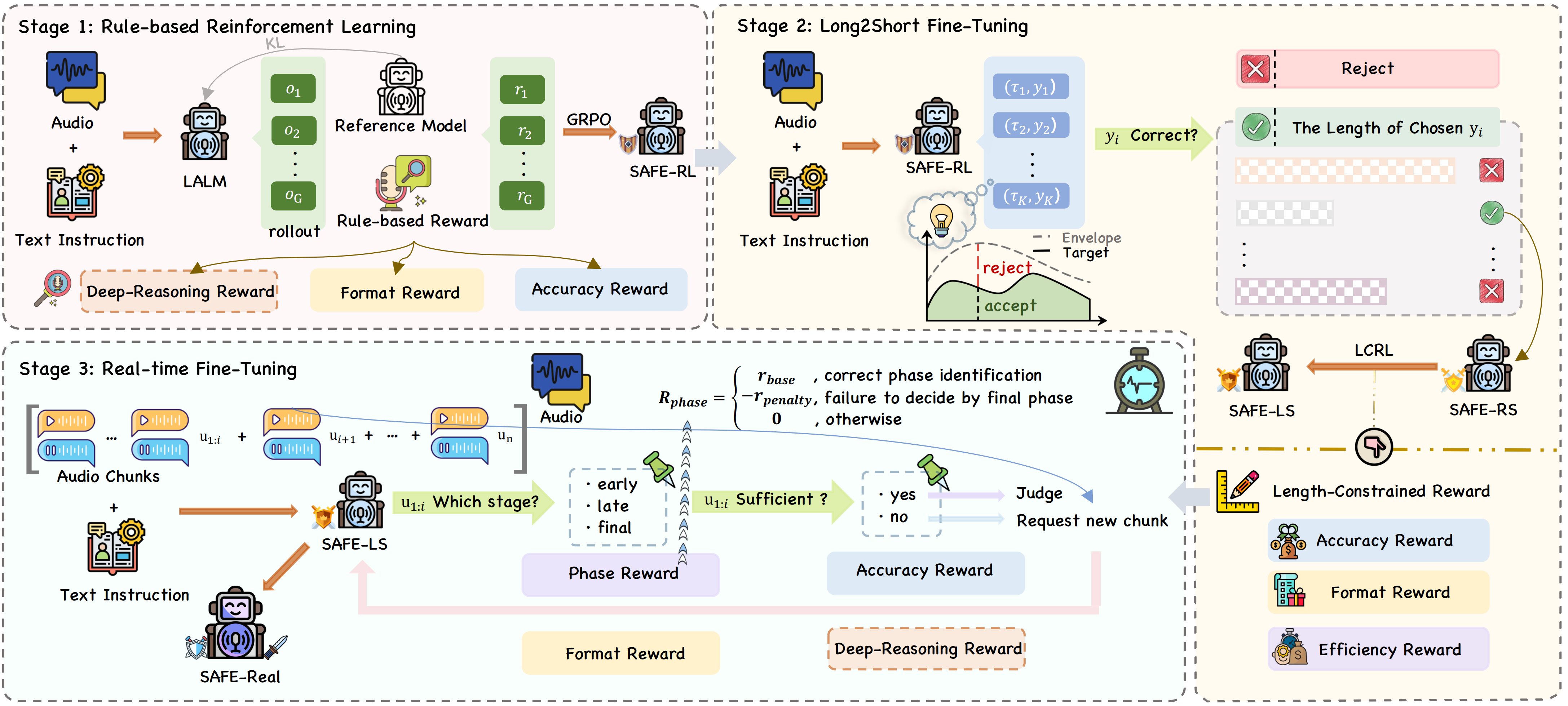}
    \caption{Overview of Our Method. Starting from an LALM, we: (i) apply rule-based RL to obtain SAFE-RL with slow-thinking capabilities; (ii) refine it using rejection sampling (SAFE-RS) and length-constrained RL (SAFE-LS) to improve reasoning efficiency; and (iii) perform real-time fine-tuning on audio chunks to derive SAFE-Real.}
    \label{fig:overview}
\end{figure*}

\subsection{Reinforcement Learning for Slow-Thinking}
Recent advancements in Reinforcement Learning (RL) have enabled LLMs to develop slow-thinking capabilities, mimicking human-like System 2 reasoning \cite{kahneman2011thinking}. Methods like OpenAI's o1/o3 \cite{jaech2024openai, OpenAIo3mini2025}, DeepSeek R1 \cite{guo2025deepseek}, and Satori \cite{shen2025satori} demonstrate notable improvements in tasks requiring step-by-step analysis, such as mathematics \cite{zhu2022solving,ludynamic}, logic \cite{jin2024impact}, and multimodal reasoning \cite{xu2024llava, thawakar2025llamav}. These models leverage techniques such as Monte Carlo Tree Search (MCTS) \cite{swiechowski2023monte} and reward-guided fine-tuning \cite{trung2024reft} to generate extended reasoning chains, enhancing their ability to solve complex problems.
However, current RL-based approaches primarily focus on text-based reasoning, leaving multimodal domains like audio-text integration underexplored. Challenges such as overthinking \cite{chen2024not} and inefficiency in dynamic scenarios \cite{qi2024interactive} highlight the need for more tailored solutions.
\section{Method}



Figure \ref{fig:overview} illustrates the three-stage framework of our approach. In Stage 1, we use rule-based reinforcement learning to train a model capable of slow-thinking. In Stage 2, we refine it via Long2Short fine-tuning to shorten reasoning and mitigate overthinking. Finally, in Stage 3, we apply Real-Time fine-tuning to optimize the model for efficient, real-time fraud detection.

\subsection{Problem Definition}

The task involves three classification objectives based on audio analysis: \textbf{scenario classification}, \textbf{fraud detection}, and \textbf{fraud type classification}. Given an input pair $(u, t)$ consisting of raw audio $u$ and text instruction $t$, the model $\pi$ generates output $o = (\tau, y)$, where $\tau$ is the step-by-step reasoning process and $y$ contains both the classification rationale and final result. The objective is to develop $\pi$ that accurately performs these tasks while providing interpretable reasoning.

\subsection{Rule-based Reinforcement Learning}
As illustrated in Figure \ref{fig:overview}, Stage 1 employs rule-based reinforcement learning for data-efficient self-evolution, yielding a slow-thinking model (SAFE-RL) that analyzes audio-text cues to detect subtle fraud patterns beyond text-only approaches.

\noindent \textbf{Group Relative Policy Optimization.} In contrast to traditional actor-critic algorithms such as Proximal Policy Optimization (PPO) \cite{schulman2017proximal}, we optimize our model using Group Relative Policy Optimization (GRPO), which eliminates the need for a critic model with parameter complexity comparable to that of the policy model $\pi_{\theta}$. Instead, GRPO estimates the relative advantage of each response based on intra-group scoring. Specifically, for each audio-text instruction pair $(u, t) \sim P(U, T)$, the policy model $\pi_{\theta}$ samples multiple reasoning processes and their corresponding responses. The output for the $i$-th sample is represented as $o_i = (\tau_i, y_i)$. Each response $y_i$ is evaluated by a rule-based reward model to compute the reward value $r_i = R(y_i)$. The intra-group relative advantage $A_i = \frac{r_i-mean(\{r1,r2,\dots,r_G\})}{std(\{r1,r2,\dots,r_G\})}$, derived from these reward values, is then used to optimize the model via the objective function $J_{GRPO}(\theta)$:

{\small
\begin{equation}
\begin{split}
&J_{GRPO}(\theta) = \mathbb{E}_{(u, t) \sim P(U,T), \{o_i\}_{i=1}^G \sim \pi_{\theta_{old}}(O|u, t)}\frac{1}{G} \\
&\sum_{i=1}^{G}(min(\rho_iA_i, clip_{\epsilon}(\rho_i)A_i)-\beta\mathbb{D}_{KL}(\pi_{\theta}\parallel\pi_{ref}))
\end{split}
\label{eq:GRPO}
\end{equation}}

The importance sampling factor $\rho_i(\theta)$ is defined as the ratio between the current policy $\pi_{\theta}$ and the sampling policy $\pi_{\theta_{\text{old}}}$. The clipping function $clip_{\epsilon}(\rho_i)$ constrains $\rho_i$ within the interval $[1 - \epsilon, 1 + \epsilon]$, ensuring conservative policy updates. The hyperparameter $\beta$ controls the strength of the KL divergence $\mathbb{D}_{KL}$ unbiasedly estimated using:

\begin{equation}
\resizebox{0.88\linewidth}{!}{$ 
\mathbb{D}_{KL}(\pi_{\theta}||\pi_{ref}) = \frac{\pi_{ref}(o_i|u,t)}{\pi_{\theta}(o_i|u,t)} - \log\frac{\pi_{ref}(o_i|u,t)}{\pi_{\theta}(o_i|u,t)} - 1
$}
\label{eq:KL}
\end{equation}

\noindent \textbf{Reward Modeling.} Training an outcome-based or process-based neural reward model is complex and resource-intensive. Transferring a general-purpose neural reward model to a specific domain requires considerable amounts of data and computational resources. In contrast, a rule-based reward model can effectively model rewards by simply designing validation rules for the answers.
\begin{itemize}
\item \textbf{Accuracy Reward ($R_{acc}$)}: Validates the final answers $y_i$ extracted from \texttt{<answer>} tags:
\begin{equation}
R_{acc} = \mathbb{I}(y_i \text{ is correct})
\label{eq:acc}
\end{equation}

\item \textbf{Format Reward ($R_{fmt}$)}: Enforces structure with \texttt{<think>} and \texttt{<answer>} tags:
\begin{equation}
R_{fmt} = \mathbb{I}(\text{format is satisfied})
\label{eq:fmt}
\end{equation}

\item \textbf{Deep-Reasoning Reward ($R_{depth}$)}: Uses length-sensitive rewards for deeper reasoning. Logarithmic normalization improves sensitivity to shorter chains:
{\small
\begin{equation}
R_{depth} =  \min\left(\frac{\ln(|\tau|+1)}{\ln(L_{max}+1)}, 1\right) \cdot R^{max}
\label{eq:depth_reward}
\end{equation}}
where $|\tau|$ is reasoning step count, $L_{max}$ the token limit and $R^{max}$ is the reward ceiling.
\end{itemize}

The total reward $R_{total}$ is computed as:
\begin{equation}
R_{total} = \alpha R_{acc} + \beta R_{fmt} + \gamma R_{depth}
\label{eq:total}
\end{equation}

We set weights $\alpha=5$ and $\beta=1$ to prioritize accuracy and format, and define $\gamma = \mathbb{I}_{\text{non-SFT}}$ to encourage deep reasoning specifically for non-SFT models. This configuration balances reliable fraud detection with structured, thorough analysis.

\subsection{Long2Short Fine-Tuning}
As shown in Figure \ref{fig:overview}, Stage 2 optimizes efficiency via \textbf{Long2Short Fine-Tuning}. This stage combines Rejection Sampling (SAFE-RS) and Length-Constrained RL (SAFE-LS) to compress reasoning chains without sacrificing precision.

\noindent \textbf{Rejection Sampling Fine-Tuning.} 
Sampling $K$ candidates $\{o_i=(\tau_i, y_i)\}_{i=1}^K$ from the proposal $\pi_\theta(\cdot|u,t)$, we define a target $\pi^*$ to prioritize correctness and brevity:

\begin{equation}
\pi^*(o|u,t) \propto \mathbb{I}(y\text{ is correct}) \cdot (1+|\tau|)^{-1}
\end{equation}
The optimal response $o^*$ is selected by maximizing:
\begin{equation}
o^* = \underset{i\in[K]}{\arg\max}\, \pi^*(o_i|u,t)
\label{eq:optimal_selection}
\end{equation}
This selects the shortest correct response. The resulting dataset $\{(u,t,o^*)\}$ is then used for SFT to train $\pi_\theta$ towards concise reasoning.

\paragraph{Length-Constrained Reinforcement Learning (LCRL).} 
Following SAFE-RS, we optimize efficiency via a composite reward $R_{LC} = \alpha R_{acc} + \beta R_{fmt} + \lambda R_{eff}$, incorporating accuracy (Eq.~\ref{eq:acc}) and format (Eq.~\ref{eq:fmt}) objectives. We set $\lambda=1$. The efficiency reward $R_{eff}$ penalizes token excess $E = \max(0, |\tau| - L_{threshold})$:
\begin{equation}
\small
R_{eff} = -\min\left(\max\left(\frac{\ln(E + 10)}{\ln(B)}, 0.1\right), 1\right) \cdot P^{max}
\label{eq:eff}
\end{equation}
where $B=1000$ controls curvature and $P^{max}$ sets the magnitude. This logarithmic scaling curbs verbosity without hindering necessary reasoning, ensuring rapid and reliable fraud detection.

\subsection{Real-Time Fine-Tuning}

Stage 3 enables dynamic risk assessment on sequential audio $u_{1:i}$. At each turn $i$, the model identifies the conversation phase (early, late, final). Through prompt engineering, we guide the model to: (1) permit early judgments, (2) formulate conclusions in late phases, and (3) mandate decisions by the final phase. We train phase awareness via:
\begin{equation}
R_{phase} = r_b \cdot \mathbb{I}_{corr} - r_p \cdot \mathbb{I}_{fail}
\end{equation}
where $\mathbb{I}_{corr}$ denotes correct phase identification, $\mathbb{I}_{fail}$ marks final phase indecision, and $r_b, r_p$ are the corresponding reward and penalty magnitudes. The total reward is $R_{total} = \alpha R_{acc} + \beta R_{fmt} + \eta R_{depth} + \delta R_{phase}$. We set $\delta=5$ and $\eta=\mathbb{I}_{\text{non-SFT}}$. This configuration balances accuracy ($R_{acc}$) with phase-appropriate decision timing ($R_{phase}$).

\begin{table*}[t]
\centering

\small
\begin{tabular}{c c c c c c  c c c c  c}
\toprule
\multirow{2}{*}{\textbf{Type}} & \multirow{2}{*}{\textbf{Model}} & \multicolumn{4}{c}{\textbf{Classification}} & \multicolumn{4}{c}{\textbf{Quality Assessment}} & \multirow{2}{*}{\textbf{Fin.}} \\
\cmidrule(lr){3-6} \cmidrule(lr){7-10}
 & & \textbf{Sce.} & \textbf{Fra.} & \textbf{FT.} & \textbf{AVG} & \textbf{Log.} & \textbf{Pra.} & \textbf{Exp.} & \textbf{SUM} & \\
\midrule
\multirow{5}{*}{ASR+LLM} 
    & GLM4-9B-Chat & 75.10 & 46.91 & 82.22 & 68.08 & 1.61 & 1.43 & 2.20 & 5.24 & 51.14 \\
    & InternLM2.5-20B & 78.34 & 36.67 & \textbf{85.42} & 66.81 & 1.99 & 1.93 & 2.43 & 6.35 & 50.21 \\
    & Qwen2.5-72B & 78.31 & \textbf{51.44} & 81.24 & \textbf{70.33} & 2.21 & 2.16 & 2.70 & 7.07 & \textbf{52.87} \\
    & Doubao 1.5 Pro & 71.14 & 36.11 & 82.25 & 63.17 & 1.94 & 1.75 & 2.60 & 6.29 & 47.48 \\
    & Deepseek V3 & \textcolor{red}{\textbf{88.53}} & 14.62 & 66.71 & 56.62 & \textbf{2.32} & \textbf{2.34} & \textbf{2.85} & \textbf{7.51} & 42.59 \\
\midrule
ASR+LRM & Deepseek R1 & 83.60 & 79.25 & 85.16 & 82.67 & \textcolor{red}{\textbf{3.18}} & \textcolor{red}{\textbf{3.26}} & \textcolor{red}{\textbf{3.50}} & \textcolor{red}{\textbf{9.94}} & 62.17 \\
\midrule
\multirow{6}{*}{LALM} 
    & GLM4-9B-Voice & 0.00 & 26.83 & 38.33 & 21.72 & 0.89 & 0.64 & 0.65 & 2.18 & 16.33 \\
    & Gemini-2-Flash & 80.51 & 59.61 & 83.53 & 74.55 & \textbf{2.25} & \textbf{2.29} & \textbf{2.72} & \textbf{7.26} & 56.03 \\
    & GPT4-o & 80.25 & 50.00 & \textbf{86.26} & 72.17 & 2.12 & 2.10 & 2.56 & 6.78 & 54.24 \\
    & Step-Audio-Chat & 76.35 & 40.65 & 79.71 & 65.57 & 1.64 & 1.62 & 2.01 & 5.27 & 49.27 \\
    & Qwen2-Audio-7B-Instruct & 70.22 & 58.51 & 20.48 & 49.74 & 1.51 & 1.42 & 1.96 & 4.89 & 37.38 \\
    & AntiFraud-Qwen2Audio & \textbf{81.31} & \textbf{84.78} & 82.91 & \textbf{83.00} & 2.06 & 2.07 & 2.31 & 6.44 & \textbf{62.36} \\
\midrule
    \cellcolor{pink!20} & \cellcolor{pink!20} \textbf{SAFE-RL} & \cellcolor{pink!20} 81.57 & \cellcolor{pink!20}\textcolor{red}{\textbf{90.20}} & \cellcolor{pink!20} \textcolor{blue}{\textbf{87.25}} & \cellcolor{pink!20} \textcolor{blue}{\textbf{86.34}} & \cellcolor{pink!20}\textcolor{blue}{\textbf{2.5}} & \cellcolor{pink!20} 2.64 & \cellcolor{pink!20} 2.97 & \cellcolor{pink!20} \textcolor{blue}{\textbf{8.11}} & \cellcolor{pink!20} \textcolor{blue}{\textbf{64.89}} \\

\cellcolor{yellow!15} LALM+Ours & \cellcolor{yellow!15} \textbf{SAFE-RS} & \cellcolor{yellow!15} 81.60 & \cellcolor{yellow!15} \textcolor{blue}{\textbf{89.61}} & \cellcolor{yellow!15} 86.39 & \cellcolor{yellow!15} 85.87 & \cellcolor{yellow!15} 2.45 & \cellcolor{yellow!15} 2.6 & \cellcolor{yellow!15}\textcolor{blue}{\textbf{2.99}} & \cellcolor{yellow!15} 8.04 & \cellcolor{yellow!15} 64.53 \\

\cellcolor{green!7} & \cellcolor{green!7} \textbf{SAFE-LS} & \cellcolor{green!7}\textcolor{blue}{\textbf{84.64}} & \cellcolor{green!7} \textcolor{blue}{\textbf{89.61}} & \cellcolor{green!7}\textcolor{red}{\textbf{88.23}} & \cellcolor{green!7}\textcolor{red}{\textbf{87.49}} & \cellcolor{green!7} 2.49 & \cellcolor{green!7}\textcolor{blue}{\textbf{2.65}} & \cellcolor{green!7} 2.97 & \cellcolor{green!7}\textcolor{blue}{\textbf{8.11}} & \cellcolor{green!7}\textcolor{red}{\textbf{65.76}}\\

\bottomrule
\end{tabular}
\caption{Performance of models on TeleAntiFraud-Bench. \textcolor{red}{\textbf{Red}} values represent SOTA results, \textcolor{blue}{\textbf{blue}} values indicate the second-best performance, and \textbf{bold} values denote the best performance within the respective model type.}
\label{tab:main}
\end{table*}

\section{Experiments}
\subsection{Experimental Setup}



\noindent \textbf{Datasets.} We utilize \textbf{TeleAntiFraud-28k} \cite{ma2025teleantifraud} (28,511 pairs) for three tasks: 7-class scenario, 2-class fraud detection, and 7-class fraud type identification. For RL, we use only raw audio and context without reasoning annotations. In Real-Time Fine-Tuning, turns $u_{1:n}$ are segmented into early ($i < n/2$), late ($n/2 \leq i < n$), and final ($i=n$) phases. Evaluation is conducted on the distribution-preserving \textbf{TeleAntiFraud-Bench} \cite{ma2025teleantifraud}.

\noindent \textbf{Baselines.} We compare our approach with proprietary models (GPT-4o \cite{hurst2024gpt}, Gemini-2-Flash \cite{GoogleDeepMindGemini2024}, Doubao-1.5 \cite{DoubaoTeamDoubao15Pro2025}) and open-source baselines including Deepseek V3/R1 \cite{liu2024deepseek, guo2025deepseek}, GLM-4-Voice \cite{zeng2024glm}, Step-Audio \cite{huang2025step}, and Qwen2-Audio variants \cite{chu2024qwen2, ma2025teleantifraud}. This selection covers reasoning specialists, ASR+LLM cascades, and end-to-end multimodal architectures across diverse scales.

\noindent \textbf{Evaluation Metrics.} We report Weighted F1 for Scenario (Sce.), Fraud (Fra.), and Type (FT.) tasks, averaged as \textbf{AVG}. Reasoning quality is scored (0-5) on Logical Rigor (Log.), Practical Value (Pra.), and Expression (Exp.), summing to \textbf{SUM} (0-15). The final metric is $\text{Fin.} = 0.75 \cdot \text{AVG} + 0.25 \cdot (\text{SUM}/15)$.

\noindent \textbf{Implementation Details.} Our backbone is AntiFraud-Qwen2Audio \cite{ma2025teleantifraud}, an SFT version of Qwen2-Audio-7B-Instruct \cite{chu2024qwen2}. The training pipeline sequentially applies Rule-based RL (SAFE-RL), Rejection Sampling (SAFE-RS), and Length-Constrained RL (SAFE-LS). Detailed hyperparameters are provided in Appendix \ref{sec:appendix_hyper}.


\subsection{Effectiveness of SAFE-QAQ in Telecom Fraud Analysis}
\noindent \textbf{Performance Improvements Across Tasks.} The experimental results in Table \ref{tab:main} reveal a consistent performance hierarchy across TeleAntiFraud-Bench tasks, with our SAFE-QAQ series models demonstrating progressive improvements over both general baselines and the specialized AntiFraud-Qwen2Audio foundation. In scenario classification ($Sce.$), while massive LLMs like Deepseek V3 (88.53) leverage their textual understanding capabilities to dominate this largely language-based task, our SAFE-LS (84.64) notably outperforms its precursor AntiFraud-Qwen2Audio (81.31) despite sharing the same architecture, confirming that our reinforcement learning framework enhances performance even for tasks where the base model already showed competence. This 3.33-point improvement is particularly notable given that the 8B-parameter AntiFraud-Qwen2Audio had already surpassed most ASR+LLM baselines through domain-specific fine-tuning.

The most striking advancements emerge in fraud detection ($Fra.$), where the evolutionary trajectory from base model to final system becomes apparent. The general-purpose Qwen2-Audio-7B-Instruct achieves only 58.51, while its SFT-enhanced version AntiFraud-Qwen2Audio reaches 84.78 through slow-thinking adaptation - already outperforming specialized text-based models like Deepseek R1 (79.25). Our SAFE-RL then extends this to 90.20 through rule-based reinforcement learning, representing a 5.42-point absolute improvement that demonstrates our method's exceptional capability in identifying subtle multimodal fraud patterns. This progression from general LALM to domain-adapted SFT model to RL-optimized system validates the complementary value of each training phase, with the final SAFE-LS achieving 88.23 in fraud-type classification ($FT.$) - surpassing even GPT4-o (86.26) and establishing new benchmarks for fine-grained multimodal analysis.

\begin{figure}[t]
    \centering
    \includegraphics[width=\linewidth]{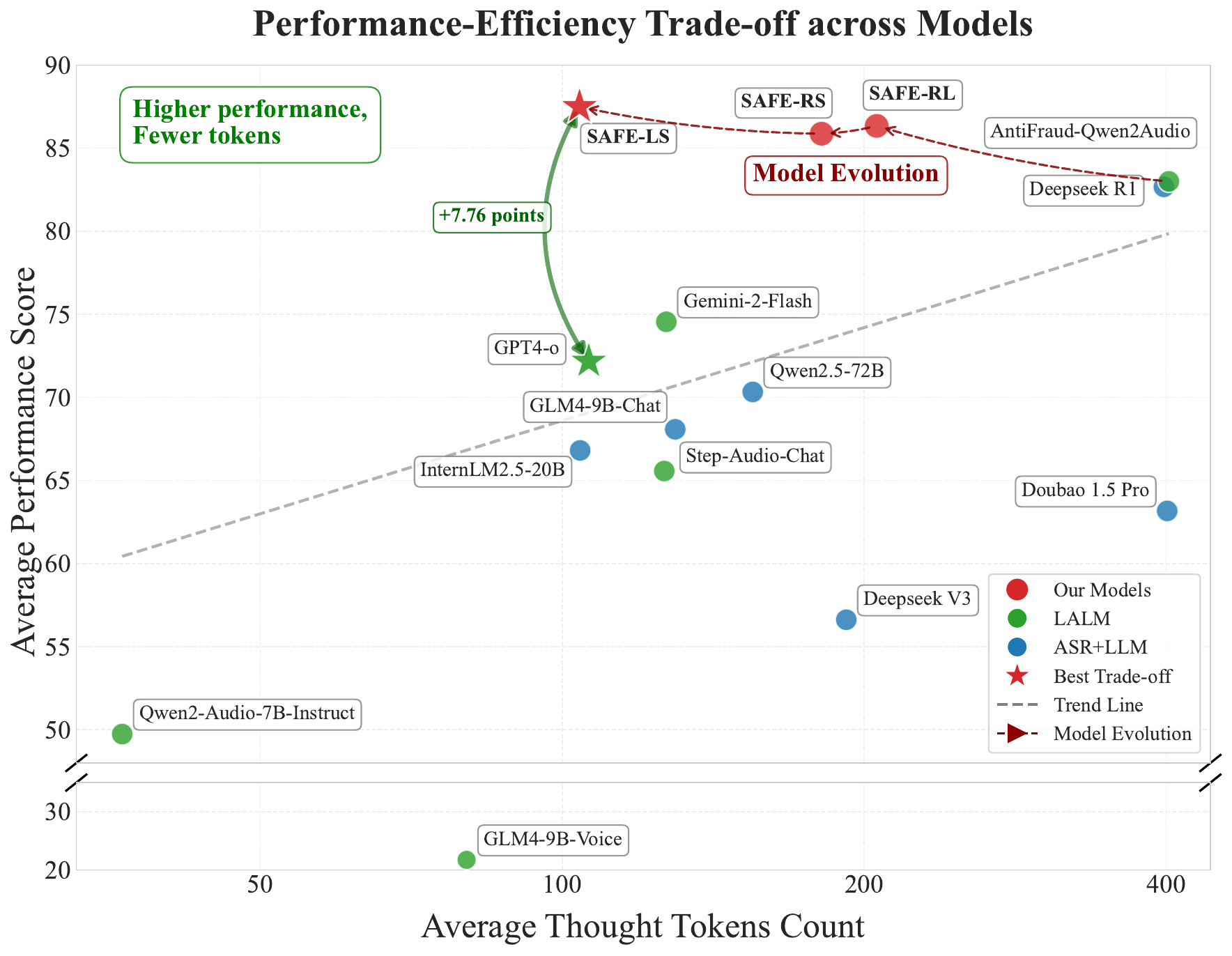}
    \caption{Performance-Efficiency Trade-off: Scatter Plot of Average Thinking Tokens vs. Average Classification Performance. Models closer to the top-left corner achieve a better balance of higher efficiency (fewer thinking tokens) and superior performance (higher classification scores). The points representing the best trade-offs for the baselines and our model are highlighted with star markers.}
    \label{fig:eff}
\end{figure}

\noindent \textbf{Reasoning Advancement.} Comprehensive metrics further validate this approach. While AntiFraud-Qwen2Audio (62.36 Fin.) already exceeds specialized text models like Deepseek R1 (62.17) through multimodal fine-tuning, our SAFE-LS (65.76) sets a new state-of-the-art through its full optimization pipeline. The 3.40-point final improvement reflects balanced advancements across all capabilities, with quality assessment scores (8.11 SUM) approaching those of dedicated reasoning models like Deepseek R1 (9.94 SUM), despite SAFE-LS using only 8B parameters compared to Deepseek R1’s 671B. This efficiency gain is achieved by our reinforcement learning framework that systematically promotes slow-thinking processes, enhancing the model's logical reasoning, practical judgment, and expressive quality. These results collectively demonstrate that while traditional approaches excel in narrow competencies, our end-to-end multimodal framework delivers superior real-world performance where detection accuracy, classification precision, and reasoning quality must operate synergistically.



\begin{table}[t]
\centering
\scriptsize
\setlength{\tabcolsep}{6pt} 
\begin{tabular}{l c c c c c c}
\toprule
\textbf{Model} & \textbf{Sce.} & \textbf{Fra.} & \textbf{FT.} & \textbf{AVG} & \textbf{Dur.} & \textbf{Turns} \\
\midrule
base & 70.22 & 58.51 & 20.48 & 49.74 & \textit{48.31s} & \textit{6.36} \\
w SFT & 81.31 & 84.78 & 82.91 & 83.00 & \textit{48.31s} & \textit{6.36} \\
\midrule
\cellcolor{pink!20}SAFE-RL & \cellcolor{pink!20}81.57 & \cellcolor{pink!20}90.20 & \cellcolor{pink!20}87.25 & \cellcolor{pink!20}86.34 & \cellcolor{pink!20}\textit{48.31s} & \cellcolor{pink!20}\textit{6.36} \\
\cellcolor{yellow!15}SAFE-RS & \cellcolor{yellow!15}81.60 & \cellcolor{yellow!15}89.61 & \cellcolor{yellow!15}86.39 & \cellcolor{yellow!15}85.87 & \cellcolor{yellow!15}\textit{48.31s} & \cellcolor{yellow!15}\textit{6.36} \\
\cellcolor{green!7}SAFE-LS & \cellcolor{green!7}84.64 & \cellcolor{green!7}89.61 & \cellcolor{green!7}88.23 & \cellcolor{green!7}87.49 & \cellcolor{green!7}\textit{48.31s} & \cellcolor{green!7}\textit{6.36} \\
\midrule
\cellcolor{gray!10}\textbf{SAFE-Real} & \cellcolor{gray!10}91.40 & \cellcolor{gray!10}88.93 & \cellcolor{gray!10}77.56 & \cellcolor{gray!10}85.96 & \cellcolor{gray!10}\textbf{8.98s} & \cellcolor{gray!10}\textbf{1.25} \\
\bottomrule
\end{tabular}
\caption{SAFE-Real vs. Baselines}
\label{tab:real_time}
\end{table}

\subsection{Performance-Efficiency Trade-off}
In real-world fraud detection systems, computational efficiency directly impacts operational costs and response time - shorter reasoning chains enable faster fraud identification during live calls while reducing infrastructure expenses. Figure \ref{fig:eff} examines the performance-efficiency trade-off across different detection systems, quantified through our proposed Thinking Efficiency Metric (TEM = $AVG/\log(|\tau|)$). This metric offers a hardware-agnostic view of efficiency based on reasoning token counts; a detailed analysis of system-level performance, including wall-clock latency and throughput, is provided in Appendix~\ref{sec:appendix_efficiency}. The scatter plot positions each model based on its average reasoning token length ($|\tau|$, log-scaled) versus average classification performance ($AVG$), revealing fundamental architectural differences and optimization trajectories. Systems positioned closer to the top-left region achieve superior balance between computational efficiency (shorter reasoning chains) and detection accuracy (higher F1 scores).

\noindent\textbf{Advantages of Audio.} LALM architectures (\textcolor{green}{green} circles) systematically outperform ASR+LLM baselines (\textcolor{blue}{blue} circles) in TEM, with average TEM scores of 33.20 versus 29.72 respectively. This efficiency advantage stems from LALMs' native multimodal processing capabilities, which not only eliminate the error accumulation inherent in cascaded ASR+LLM pipelines but also capture audio-specific semantic information beyond just speech content, encompassing rich paralinguistic signals.

\noindent\textbf{Scaling Laws in Fraud Reasoning.} we observe a scaling law-like relationship between reasoning complexity and performance gains, where increasing the reasoning tokens yields logarithmic improvements in detection accuracy. Our analysis reveals remarkably consistent scaling patterns: when comparing GPT4-o to Gemini-2-Flash ($\Delta log(|\tau|)=0.0773$) and GLM4-9B-Chat to Qwen2.5-72B ($\Delta log(|\tau|)=0.0772$), we find nearly identical performance gains ($\Delta AVG=2.38$ vs $2.25$ respectively), with the ratio $\frac{\Delta log(|\tau|)}{\Delta AVG}$ remaining stable across model families (0.0325±0.0018). This scaling behavior suggests that fraud reasoning tasks exhibit fundamental dynamics similar to those observed in large language model pre-training, though our reinforcement learning framework ultimately breaks this pattern through targeted optimization.

\begin{table}[t]
\centering
\small
\begin{tabular}{l c c c c}
\toprule
\textbf{Model} & \textbf{Sce.} & \textbf{Fra.} & \textbf{FT.} & \textbf{AVG} \\
\midrule
\cellcolor{pink!20}\textbf{SAFE-RL} & \cellcolor{pink!20}81.57 & \cellcolor{pink!20}90.20 & \cellcolor{pink!20}87.25 & \cellcolor{pink!20}\textbf{86.34} \\
w/o SFT & 73.37 & 85.70 & 82.91 & 80.66 \\
w/o SFT w $R_{\text{depth}}$ & 77.25 & 86.97 & 81.94 & 82.05 \\
\midrule
\cellcolor{yellow!15}\textbf{SAFE-RS} & \cellcolor{yellow!15}81.60 & \cellcolor{yellow!15}89.61 & \cellcolor{yellow!15}86.39 & \cellcolor{yellow!15}\textbf{85.87} \\
w/o SFT & 79.81 & 87.14 & 83.82 & 83.59 \\
\midrule
\cellcolor{green!7}\textbf{SAFE-LS} & \cellcolor{green!7}84.64 & \cellcolor{green!7}89.61 & \cellcolor{green!7}88.23 & \cellcolor{green!7}\textbf{87.49} \\
w/o SFT & 82.89 & 91.42 & 87.07 & 87.13 \\
w/o RS & 82.39 & 90.43 & 86.10 & 86.31 \\
\bottomrule
\end{tabular}
\caption{Ablation Study: Performance of Our Models}
\label{tab:abl1}
\end{table}

\noindent\textbf{Effectiveness of Multi-Stage Optimization.} Our SAFE optimization pathway (\textcolor[RGB]{139,0,0}{dark red} trajectory) demonstrates systematic efficiency gains while maintaining performance superiority. Our optimization starts from the slow-thinking AntiFraud-Qwen2Audio (TEM=31.87), rule-based reinforcement learning in SAFE-RL reduces average reasoning tokens by 48.87\% while improving $F1_{avg}$ by 3.34 points (TEM=37.32). Subsequent rejection sampling fine-tuning (SAFE-RS) achieves additional 11.87\% token reduction before length-constrained RL finalizes the optimization in SAFE-LS (TEM=43.38). This three-stage refinement yields 36.12\% higher TEM than the original base model, ultimately outperforming GPT4-o's TEM by 7.76 points through coordinated reasoning compression and performance enhancement.

\noindent\textbf{Cost-Effective SOTA Performance.} The rightmost cluster contains specialized reasoning models like Deepseek R1 (TEM=16.03) that use exhaustive token generation (>397 tokens on average) to achieve competitive accuracy. While these systems approach the performance ceiling, their operational costs hinder real-world deployment. 
Our SAFE-LS achieves state-of-the-art accuracy (88.23 $AVG$) using only 25.85\% of the tokens required by comparable models. This efficiency enables our deployed system to process over 70,000 screened calls daily, significantly reducing manual audit burdens and preventing financial losses through automated intervention. This sets a new practical benchmark for fraud detection.

\begin{figure*}[th]
    \centering
    \includegraphics[width=1.0\linewidth]{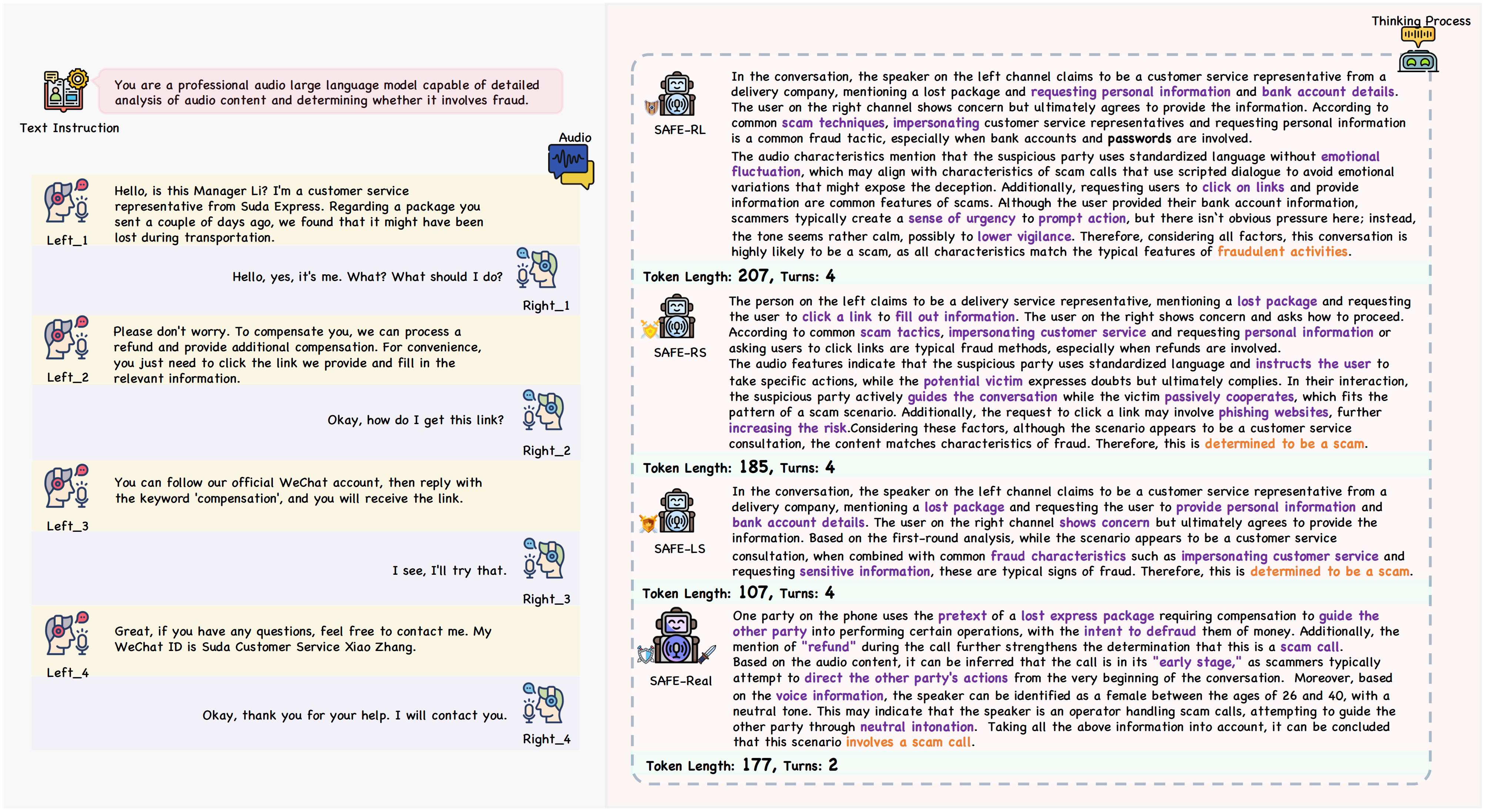}
    \caption{Model Output Case Study: Input with Text Instruction and Audio (ASR Results for Clarity, Left), Reasoning Process of SAFE-QAQ Series (Right). Key reasoning points are highlighted in \textcolor{violet}{purple}, and inference results are marked in \textcolor{orange}{orange}.}
    \label{fig:case}
\end{figure*}

\begin{table}[t]
\centering
\small
\begin{tabular}{l c c c c c}
\toprule
\textbf{Model} & \textbf{Log.} & \textbf{Pra.} & \textbf{Exp.} & \textbf{Sum} & \textbf{tokens} \\
\midrule
\cellcolor{pink!20}\textbf{SAFE-RL} & \cellcolor{pink!20}2.50 & \cellcolor{pink!20}2.64 & \cellcolor{pink!20}2.97 & \cellcolor{pink!20}8.11 & \cellcolor{pink!20}205.76 \\
w/o SFT & 1.93 & 1.92 & 2.39 & 6.24 & 31.46 \\
w/o SFT w $R_{\text{depth}}$ & 2.10 & 2.07 & 2.59 & 6.76 & 212.72 \\
\midrule
\cellcolor{yellow!15}\textbf{SAFE-RS} & \cellcolor{yellow!15}2.45 & \cellcolor{yellow!15}2.60 & \cellcolor{yellow!15}2.99 & \cellcolor{yellow!15}8.04 & \cellcolor{yellow!15}181.33 \\
w/o SFT & 2.09 & 2.09 & 2.60 & 6.78 & 173.78 \\
\midrule
\cellcolor{green!7}\textbf{SAFE-LS} & \cellcolor{green!7}2.49 & \cellcolor{green!7}2.65 & \cellcolor{green!7}2.97 & \cellcolor{green!7}8.11 & \cellcolor{green!7}104.02 \\
w/o SFT & 2.17 & 2.25 & 2.64 & 7.06 & 74.05 \\
w/o RS & 2.23 & 2.31 & 2.78 & 7.32 & 106.20 \\
\bottomrule
\end{tabular}
\caption{Ablation Study: Reasoning Quality and Token Efficiency of Our Models}
\label{tab:abl2}
\end{table}

\subsection{Real-Time Detection Performance}  

Table \ref{tab:real_time} demonstrates that SAFE-Real achieves real-time detection with an average duration of just 8.98 seconds (81.4\% faster than non-real-time models), while maintaining robust fraud detection performance (88.93 F1). Although its fraud-type classification accuracy decreases by 12.2\% (77.56 F1) compared to SAFE-LS, this trade-off is operationally justified: in real-world fraud prevention, early detection takes precedence over precise typing, as promptly stopping active scams can prevent financial losses. The model's superior scenario classification (91.40 F1) and ultra-low average of 1.25 conversational turns enable highly effective live call interception. 

\subsection{Ablation Studies}


The ablation studies in Tables \ref{tab:abl1} and \ref{tab:abl2} demonstrate the critical role of SFT pretraining and the progressive improvements of our SAFE framework. Removing SFT leads to notable degradation across all metrics (e.g., SAFE-RL's performance drops 5.68 points to 80.66 and reasoning quality decreases 1.87 points to 6.24), though incorporating Deep-Reasoning Reward ($R_{\text{depth}}$) without SFT partially mitigates these losses (improving to 82.05 and 6.76 respectively). Our token-efficient variants (SAFE-RS/SAFE-LS) maintain strong performance (85.87/87.49) while using substantially fewer tokens (181.33/104.02 vs 205.76), while preserving high reasoning quality (8.04/8.11 respectively). Yet they still benefit from SFT's foundational capabilities: Removing SFT causes SAFE-RS/SAFE-LS to drop to 83.59/87.13 with lower reasoning quality (6.78/7.06). This confirms SFT's essential role in establishing baseline abilities that subsequent RL stages enhance rather than replace. Notably, when we remove rejection sampling in the Long2Short stage (SAFE-LS w/o RS), we observe performance degradation (87.49 to 86.31) along with decreased reasoning quality, demonstrating the necessity of Rejection Sampling Fine-Tuning for maintaining performance while improving efficiency in subsequent stages.

\section{Case Study}
Figure \ref{fig:case} shows our models' reasoning processes in a typical "lost package refund" scam scenario. From SAFE-RL to SAFE-RS and SAFE-LS, the length of the models' reasoning processes progressively shortens (from 207 to 107 tokens) as a result of the Long2Short optimization. Concurrently, the density of key reasoning points increases, demonstrating that Long2Short enables more efficient reasoning and enhances model efficiency. Due to the analysis to assess the current stage of the call, SAFE-Real employs a moderate-length reasoning process. Notably, SAFE-Real achieves interpretable fraud detection using only two rounds of dialogue audio, underscoring its high efficiency. Critically, our models extract essential paralinguistic cues (accents, emotions, vocal tones) directly from raw audio, which ASR transcriptions would lose, thereby exposing the subtle mismatch between calm tone and urgent intent.

\section{Conclusion}




We present SAFE-QAQ, an end-to-end slow-thinking audio-text fraud detection framework trained via reinforcement learning. By integrating GRPO with rule-based rewards, Long2Short optimization, and real-time learning, our model achieves state-of-the-art performance on TeleAntiFraud-Bench (88.23 F1) with significantly improved efficiency (48.87\% shorter chains and 81.4\% faster speed). Beyond academic metrics, SAFE-QAQ has been successfully deployed in a production pipeline processing over 70,000 calls daily, where it effectively reduces manual audit burdens and prevents financial losses. This work validates that multimodal slow-thinking architectures can be both robust and practically efficient, offering a scalable solution for real-world security challenges.

\section*{Limitation}
While SAFE-QAQ demonstrates superior performance, our experimental scope is inevitably constrained by the fact that TeleAntiFraud-28k is currently the only open-source dataset suitable for training Large Audio-Language Models for fraud detection. This data scarcity restricts a more extensive evaluation of generalization capabilities across highly diverse fraud scenarios or unexpected acoustic conditions. Consequently, although our model shows strong noise resilience, further validation is required to ensure robustness in extreme real-world environments with severe interference or signal degradation.

\section*{Ethical Statement}
This research upholds strict data privacy standards. For experimental validation, we used the anonymized TeleAntiFraud-28k dataset, ensuring no exposure of personally identifiable information (PII). Regarding the real-world deployment, our system is implemented in collaboration with telecom operators via privacy-preserving intermediate number services. This approach masks actual phone numbers, with all data processing authorized by enterprises strictly for anti-fraud quality inspection. We emphasize that SAFE-QAQ is designed exclusively as a defensive tool, and we mandate continuous monitoring to prevent misuse and mitigate potential algorithmic bias in practical applications.

\bibliography{custom}
\newpage
\appendix


\section{Computational Efficiency Analysis}
\label{sec:appendix_efficiency}

To validate that our theoretical reductions in reasoning token counts translate into practical wall-clock time savings, we conducted a detailed profiling of inference latency and throughput. 

\paragraph{Experimental Setup.} All efficiency evaluations were performed on NVIDIA A100 GPUs. We measured the performance of our three iterative models: SAFE-RL, SAFE-RS, and SAFE-LS. To ensure consistent and reproducible measurements, we utilized greedy decoding (temperature $= 0$). We report the median latency ($p50$), the 95th percentile latency ($p95$), and the overall system throughput (samples per second).

\paragraph{Results.} As presented in Table~\ref{tab:latency_profile}, the optimization stages demonstrate a clear trajectory of efficiency improvement. SAFE-RL, which produces the longest reasoning chains, exhibits the highest latency. The Rejection Sampling stage (SAFE-RS) provides a moderate improvement by filtering out inherently long-winded responses. Most notably, the final Length-Constrained RL stage (SAFE-LS) achieves a $p50$ latency of 916.2ms and a throughput of 1.10 samples/s.

This corresponds to a $\approx 26.34\%$ reduction in median latency compared to the SAFE-RL baseline. These results confirm that the "Thinking Efficiency Metric" (TEM) discussed in Section 4.3 correlates strongly with real-world deployment metrics. The substantial drop in $p95$ latency (from 1895.7ms to 1207.8ms) also indicates that SAFE-LS is significantly more stable and robust against "infinite loops" or excessive overthinking, making it more suitable for time-sensitive fraud detection scenarios.

\section{Prompts}

\begin{figure*}[th]
    \centering
    \includegraphics[width=1.0\linewidth]{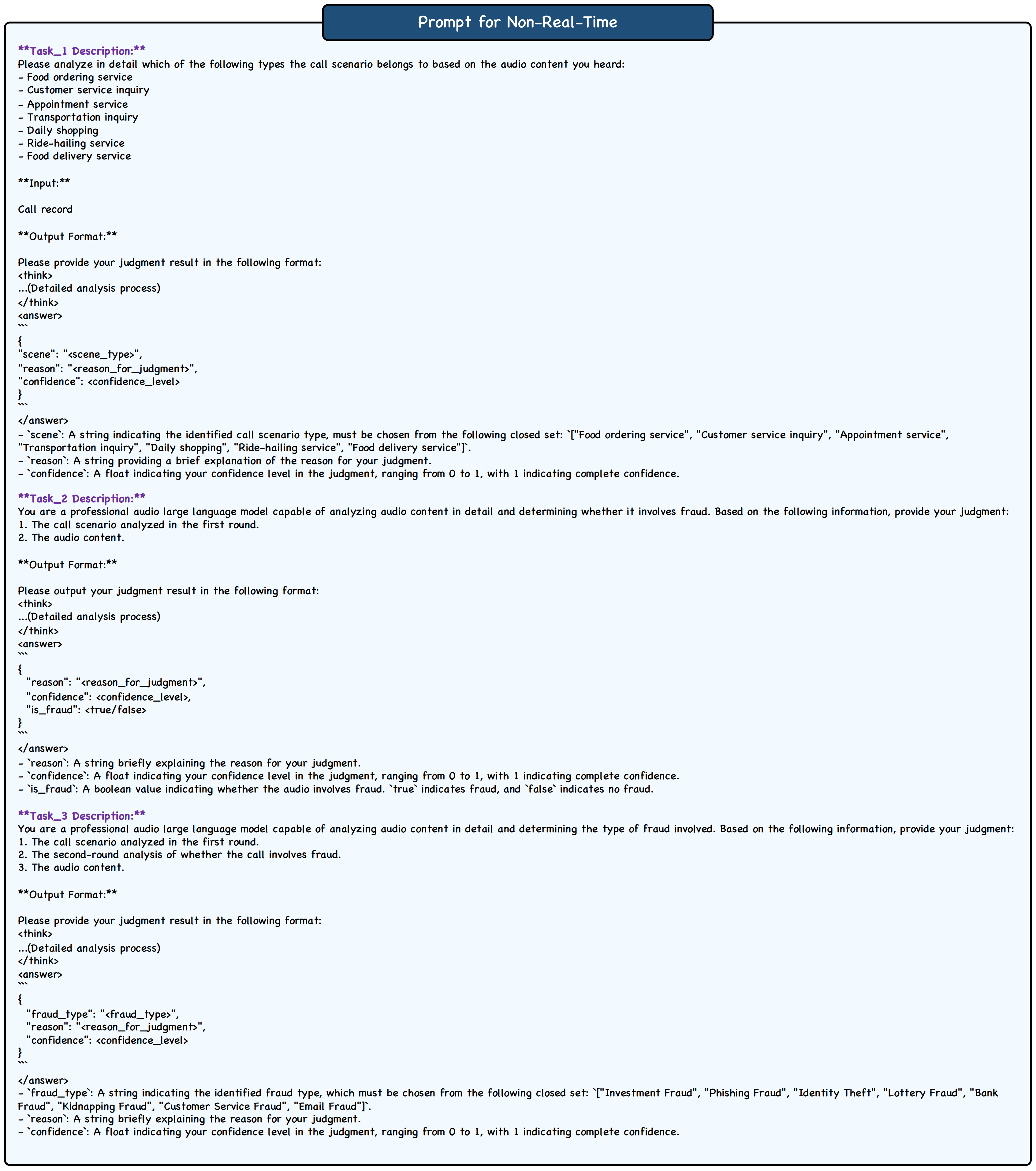}
    \caption{Prompt for Non-Real-Time.}
    \label{fig:prompt1}
\end{figure*}

\begin{figure*}[th]
    \centering
    \includegraphics[width=1.0\linewidth]{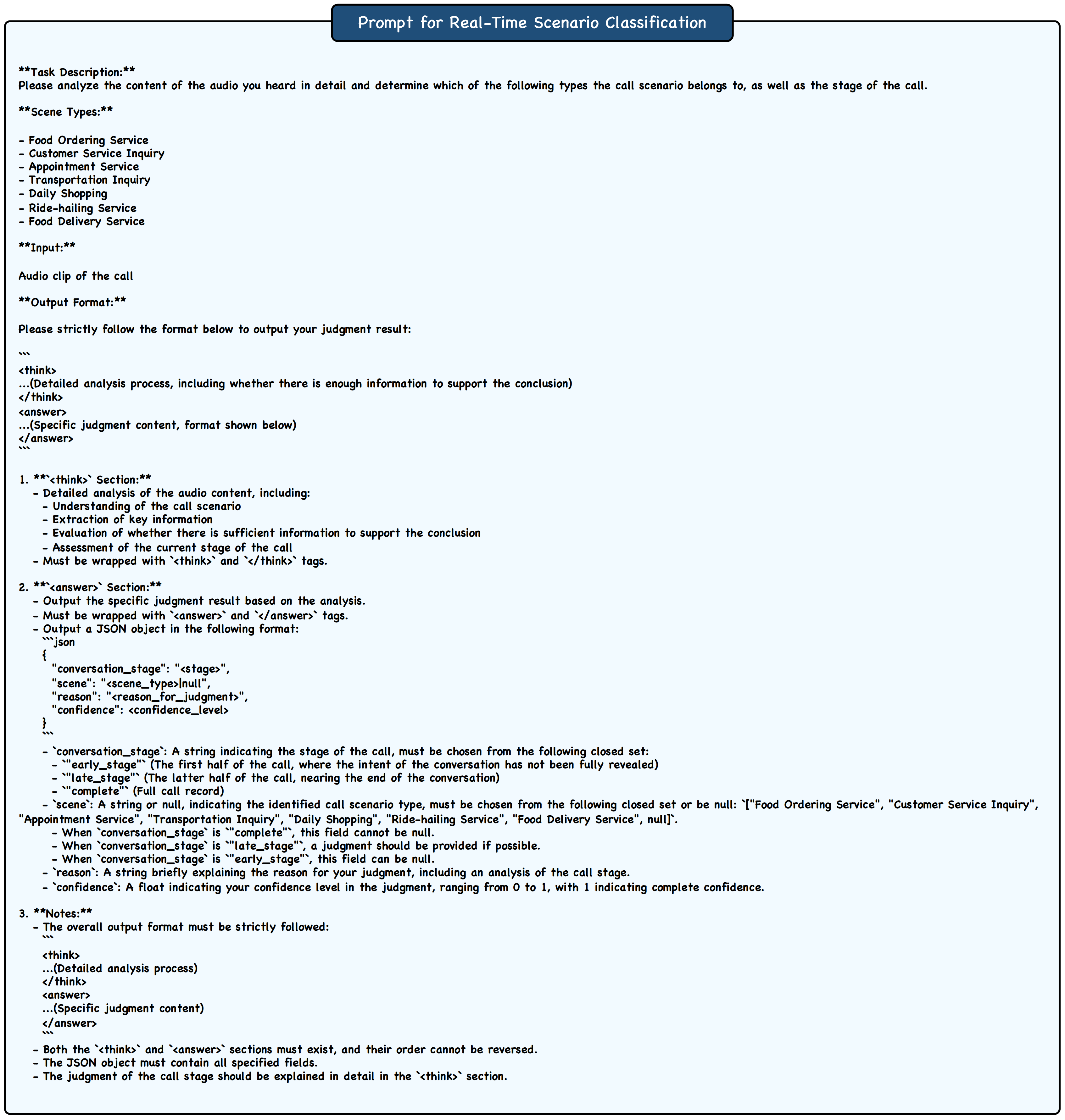}
    \caption{Prompt for Real-Time Scenario Classification.}
    \label{fig:prompt2_1}
\end{figure*}

\begin{figure*}[th]
    \centering
    \includegraphics[width=1.0\linewidth]{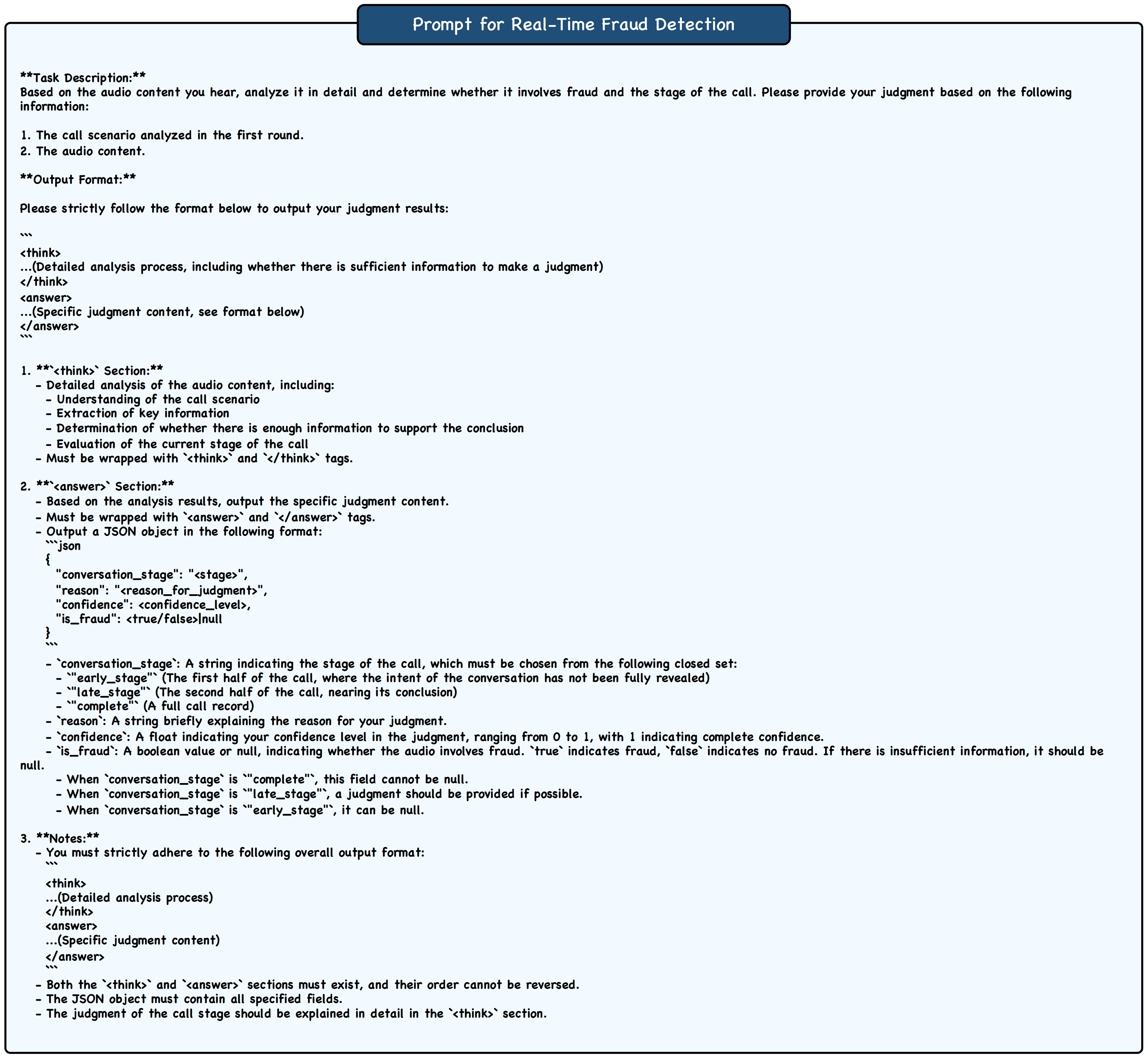}
    \caption{Prompt for Real-Time Fraud Detection.}
    \label{fig:prompt2_2}
\end{figure*}

\begin{figure*}[th]
    \centering
    \includegraphics[width=1.0\linewidth]{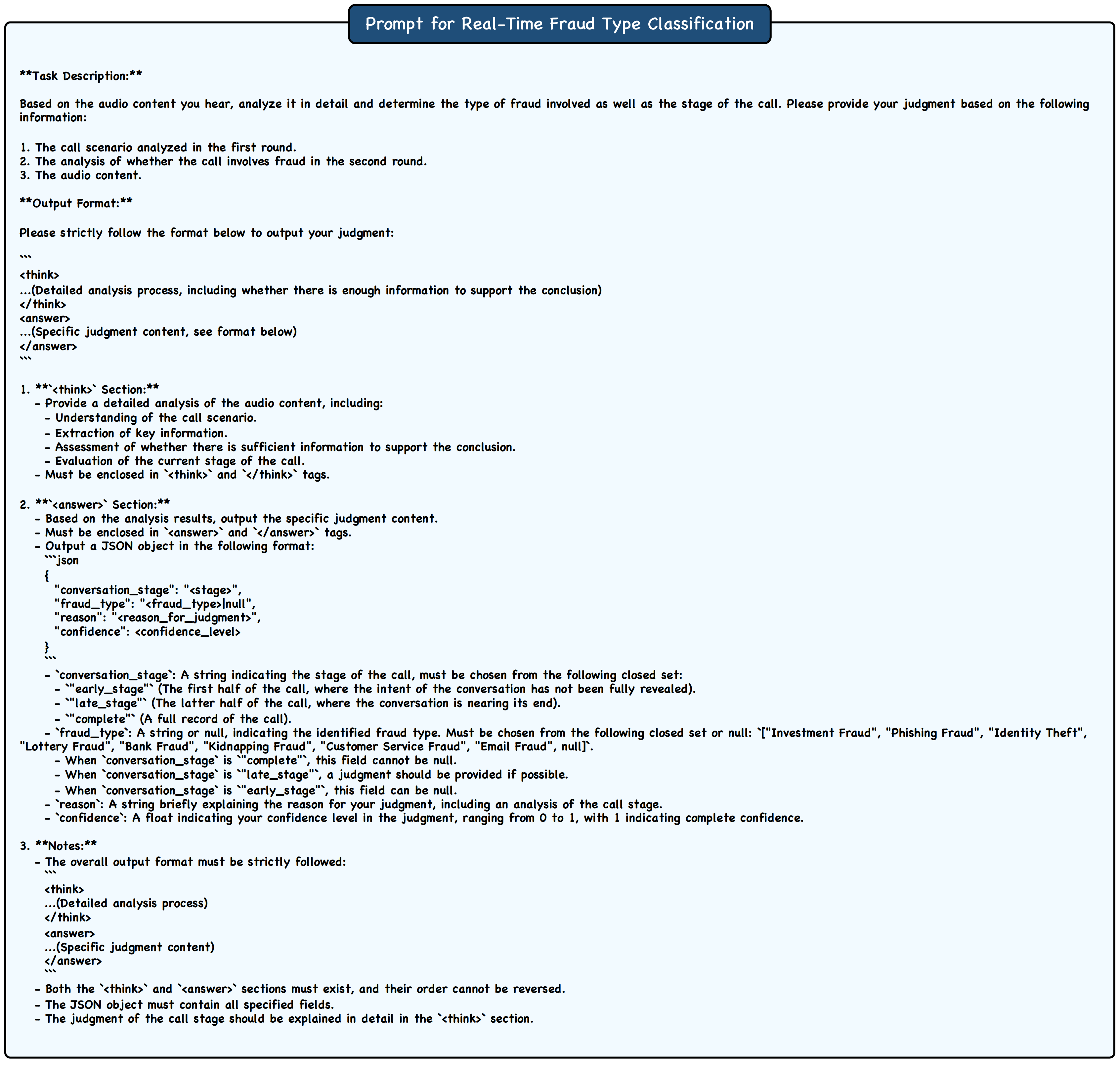}
    \caption{Prompt for Real-Time Fraud Type Classification.}
    \label{fig:prompt2_3}
\end{figure*}

\begin{figure*}[th]
    \centering
    \includegraphics[width=1.0\linewidth]{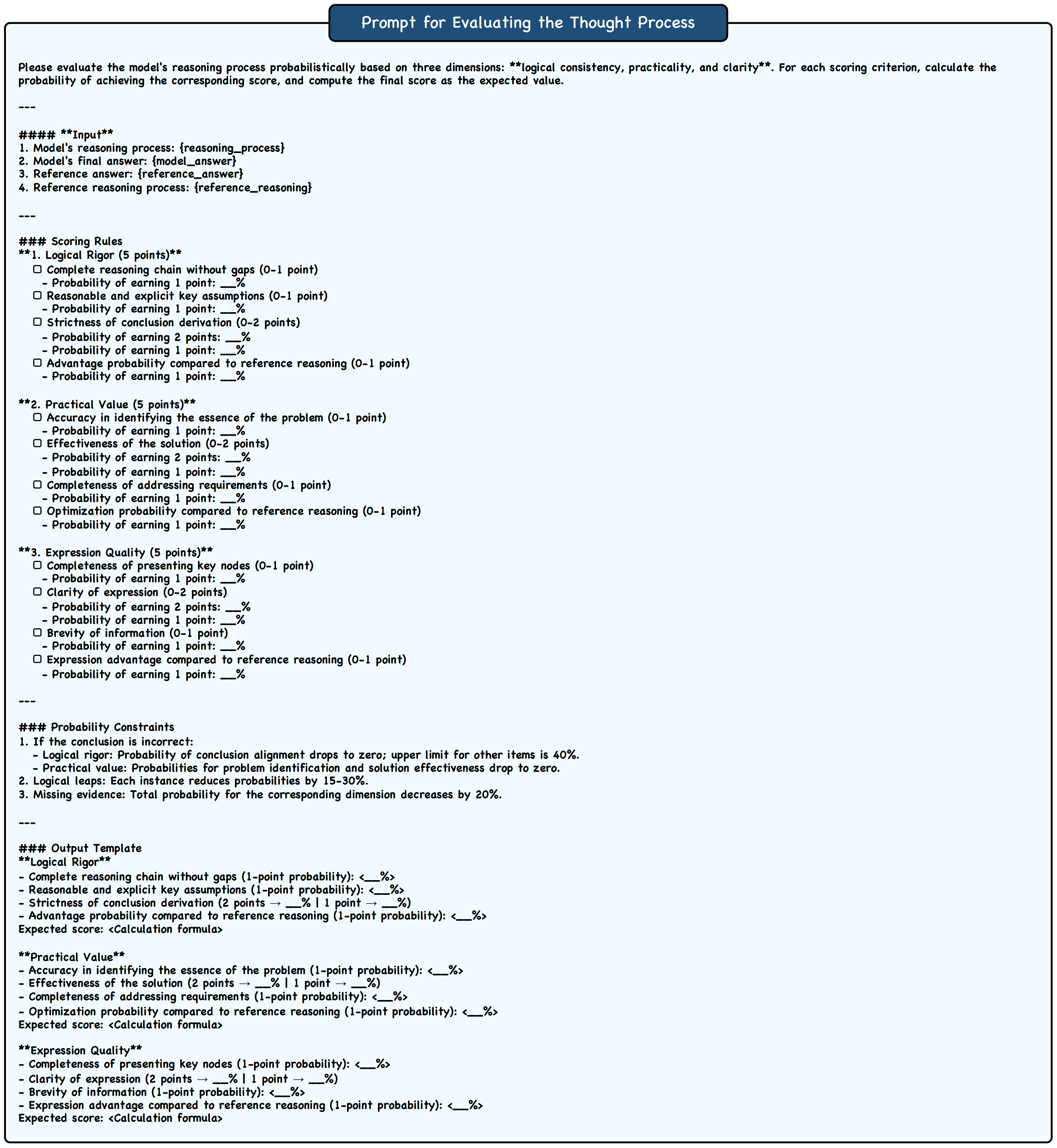}
    \caption{Prompt for Evaluating the Thought Process.}
    \label{fig:prompt3}
\end{figure*}

The prompts used in this study are designed to guide the model through various stages of reasoning and decision-making. Below is a detailed description of their roles:
\begin{itemize}
 \item Figure \ref{fig:prompt1} illustrates the prompts utilized in the first and second stages for training across three tasks: scenario classification, fraud detection, and fraud type classification. These prompts are structured to facilitate slow-thinking reasoning, enabling the model to capture subtle discrepancies in audio details, such as vocal tone fluctuations, emotional stress, and environmental cues.

 \item Figure \ref{fig:prompt2_1}, Figure \ref{fig:prompt2_2}, and Figure \ref{fig:prompt2_3} present the prompts employed during the real-time detection phase. These prompts dynamically adjust based on the conversation phase (early, late, or final) to ensure timely and accurate fraud detection while considering the sufficiency of available information.

 \item Figure \ref{fig:prompt3} showcases the prompts designed for evaluating the quality of the model’s reasoning process. These prompts focus on assessing logical rigor, practical value, and expression quality, providing a comprehensive evaluation framework for the model’s performance.
\end{itemize}

\begin{table}[t]
    \centering
    \caption{Inference Latency and Throughput Profiling on NVIDIA A100. comparison across the three training stages shows that our Long2Short optimization (SAFE-LS) significantly reduces latency and improves throughput.}
    \label{tab:latency_profile}
    \resizebox{0.99\linewidth}{!}{
    \begin{tabular}{lccc}
        \toprule
        \textbf{Model} & \textbf{p50} & \textbf{p95} & \textbf{Throughput} \\
        \midrule
        SAFE-RL & 1243.9 & 1895.7 & 0.76 \\
        SAFE-RS & 1204.0 & 1263.4 & 0.85 \\
        SAFE-LS & \textbf{916.2} & \textbf{1207.8} & \textbf{1.10} \\
        \bottomrule
    \end{tabular}
    }
\end{table}

\section{Hyperparameter Settings and Sensitivity Analysis}
\label{sec:appendix_hyper}

To ensure reproducibility and facilitate further research, we provide a comprehensive detailed description of our hyperparameter configurations, including the rationale behind specific choices and sensitivity analyses conducted during the development of SAFE-QAQ.

\subsection{Implementation Platform}
All experiments were conducted on a high-performance computing cluster equipped with 4 NVIDIA A100 (80GB) GPUs. We utilized a global batch size of 12 (implemented as a per-device batch size of $bs=3$ with gradient accumulation). The training framework is built upon \texttt{ms-swift}\footnote{\url{https://github.com/modelscope/ms-swift}}, optimized for efficient large audio-language model fine-tuning.

\subsection{Configuration Rationale}
Our hyperparameter selection strategy balances training stability, inference efficiency, and task performance. The specific configurations are categorized as follows:

\noindent \textbf{Reward Weight Configuration.} 
The composite reward functions involve multiple components ($\alpha, \beta, \delta, \lambda$). We determined their values based on the priority of objectives:
\begin{itemize}
    \item \textbf{Accuracy Priority ($\alpha=5, \beta=1$):} We set the ratio $\alpha:\beta=5:1$. This heavy weighting on $\alpha$ ensures that the model prioritizes the correctness of the final classification (Accuracy Reward) over mere structural compliance (Format Reward), while $\beta=1$ remains sufficient to guide the parser.
    \item \textbf{Phase Awareness ($\delta=5$):} For Real-Time Fine-Tuning, accurate phase recognition is critical for timely intervention. We set $\delta=5$, equal to the accuracy weight $\alpha$, to emphasize that identifying the correct conversation phase is as substantial as the fraud detection itself in live scenarios.
    \item \textbf{Efficiency Balance ($\lambda=1$):} We set $\lambda=1$ to introduce a regularization term for reasoning length. This value was chosen to curb verbosity without overpowering the accuracy reward, preventing the model from sacrificing necessary reasoning depth for brevity.
\end{itemize}

\noindent \textbf{Reinforcement Learning (GRPO) Parameters.}
We adopt the Group Relative Policy Optimization (GRPO) algorithm.
\begin{itemize}
    \item \textbf{Stability Factors ($\epsilon=0.2, \beta_{KL}=0.04$):} We adhere to the default settings recommended by the \texttt{ms-swift} framework. Specifically, the clipping coefficient $\epsilon=0.2$ and the KL divergence coefficient $\beta_{KL}=0.04$ are crucial for preventing policy collapse. In our preliminary experiments, we explored removing the KL penalty (i.e., $\beta_{KL}=0$), which resulted in severe training instability and mode collapse. Thus, we retained the robust default values.
    \item \textbf{Group Size ($G=9$):} We set the group size to 9. This value represents a trade-off between computational overhead and gradient variance reduction, ensuring stable convergence within limited GPU memory.
\end{itemize}

\noindent \textbf{Generation and Length Constraints.}
\begin{itemize}
    \item \textbf{Sampling Strategy:} To balance generation diversity and quality during exploration, we utilize Nucleus Sampling with $top\_p=0.9$, $top\_k=50$, and a temperature of $0.9$. For Rejection Sampling, we set the number of candidates $K=16$ to ensure sufficient coverage of the solution space.
    \item \textbf{Length Thresholds:} The maximum length threshold $L_{max}=200$ and threshold $L_{threshold}=200$ for $R_{depth}$ and $R_{eff}$ were determined based on the statistical distribution of reasoning chains in the TeleAntiFraud-28k dataset. $P^{max}=5$ is set to cap the penalty magnitude, preventing excessive gradients that could destabilize the policy.
\end{itemize}

\subsection{Hyperparameter Sensitivity Analysis}
To validate our choice of learning rate, which is a critical factor in RL convergence, we conducted a grid search using the SAFE-RL (w/o SFT) model on a synthetic subset of TeleAntiFraud-Bench. The results are summarized in Table \ref{tab:lr_ablation}.

\begin{table}[h]
\centering
\caption{Sensitivity analysis of Learning Rate (LR) on model performance. The selected setting ($3e^{-5}$) achieves the best balance across all metrics.}
\label{tab:lr_ablation}
\setlength{\tabcolsep}{3.5mm}
\resizebox{0.99\linewidth}{!}{\begin{tabular}{l|cccc}
\toprule
\textbf{Learning Rate} & \textbf{Sce.} & \textbf{Fra.} & \textbf{FT.} & \textbf{AVG} \\
\midrule
$1e^{-5}$ & 85.34 & 72.86 & 75.36 & 77.85 \\
$\mathbf{3e^{-5}}$ \textbf{(Ours)} & 84.31 & \textbf{88.70} & 75.74 & \textbf{82.92} \\
$5e^{-5}$ & 85.12 & 78.10 & \textbf{76.56} & 79.93 \\
\bottomrule
\end{tabular}}
\end{table}

As observed, a learning rate of $3e^{-5}$ yields the highest average F1 score (AVG: 82.92). Lower rates ($1e^{-5}$) resulted in underfitting, particularly in the Fraud Detection (Fra.) task, while higher rates ($5e^{-5}$) degraded performance, likely due to optimization overshooting. Consequently, $lr=3e^{-5}$ was selected for all main experiments.

\end{document}